\documentclass[conference,a4]{IEEEtran}
\IEEEoverridecommandlockouts
\usepackage{cite}
\usepackage{amsmath,amssymb,amsfonts}
\usepackage{algorithmic}
\usepackage{graphicx}
\usepackage{textcomp}
\usepackage{xcolor}
\usepackage{color}
\usepackage{bm}
\usepackage{latexsym}
\def\qed{\hfill $\Box$}
\newtheorem{theorem}{Theorem}

\begin{document}

\title{Precoder Design for Correlated Data Aggregation 
\\via Over-the-Air Computation in Sensor Networks
\thanks{This work was supported by JSPS Grant-in-Aid for Scientific Research(A) Grant Number JP22H00514.}
}

\author{\IEEEauthorblockN{Ayano Nakai-Kasai}
\IEEEauthorblockA{\textit{Graduate School of Engineering} \\
\textit{Nagoya Institute of Technology}\\
Nagoya, Japan \\
nakai.ayano@nitech.ac.jp}
\and
\IEEEauthorblockN{Tadashi Wadayama}
\IEEEauthorblockA{\textit{Graduate School of Engineering} \\
\textit{Nagoya Institute of Technology}\\
Nagoya, Japan \\
wadayama@nitech.ac.jp}
}

\maketitle

\begin{abstract} 
    Over-the-air computation (AirComp) enables efficient wireless data aggregation in sensor networks 
    by simultaneous processing of calculation and communication.
	This paper proposes a novel precoder design for AirComp 
    that incorporates statistical properties of sensing data, spatial correlation and heterogeneous data correlation.
	The proposed design of the precoder requires no iterative processes 
    so that it can be realized with low computational costs.
	Moreover, this method provides dimensionality reduction of sensing data 
    to reduce communication costs per sensor.
	We evaluate performance of the proposed method in terms of various system parameters.
	The results show the superiority of the proposed method to conventional non-iterative methods 
	in cases where there are a large number of sensors and 
    where the number of receive antennas at the aggregator is less than 
	that of the total transmit antennas at the sensors.
\end{abstract}


\section{Introduction}
In the fifth or more generation communication systems, 
one of the core technologies is to connect a large number of Internet-of-Things (IoT) devices 
that have abilities of sensing, computation, and wireless communication 
and to utilize their sensing data for many practical applications \cite{Salam}, \cite{Zhu2021}.
There are a lot of active applications of sensor networks composed of the sensing devices, 
such as in agriculture \cite{Elijah} and in environmental monitoring \cite{Smidl}, \cite{Othman}.
In a centralized data processing, 
data from distributed sensing devices are collected via wireless communication 
at an aggregator, which performs calculations 
to achieve desired actions for the applications. 
This procedure is called wireless data aggregation.
It is desirable to achieve the wireless data aggregation with low latency 
for immediate response to demands in large-scale IoT networks.

The idea of over-the-air computation (AirComp) was first investigated in the field of information theory \cite{Nazer} 
and has been gathering much attention from a signal processing perspective 
in sensor networks and wireless data aggregation \cite{Zhu2021}, \cite{Goldenbaum}.
AirComp enables fast wireless data aggregation 
by jointly receiving transmitted signals and calculating some function value of the signals. 
This is achieved by simultaneous transmission at sensor nodes over the same frequency band 
and by obtaining the sum of the transmitted signals with the analog-wave superposition property of wireless multiple-access channels (MAC).
It is different from classical data aggregation settings such as time-division based schemes 
where all the data are separately received 
and then the function value is calculated at the aggregator. 
In AirComp, efficient processes of sensing data can be achieved with low latency 
and the required bandwidth does not depend on the number of sensor nodes 
so that it is suitable for large-scale IoT networks.

AirComp in sensor networks occurs aggregation errors, 
i.e.,  errors between the actual sum of transmitted signals from the sensors and the aggregated data on the air, 
due to different channel coefficients of the sensors and additive noise at the aggregator.
To reduce the aggregation errors, 
scaling coefficients or matrices are applied as precoders to the transmit signals for the sensors.
The optimization of them has been tackled in various contexts \cite{Liu, WLiu, Cao, Zhu, Huh, Chen}.
% In addition, these methods do not consider correlation among the sensors 
% which is potentially useful for improving the system performance.

For designing an efficient precoder for AirComp, 
channel state information, transmit power constraints, and statistical properties of data at the sensors 
should be taken into consideration. 
Especially in sensor networks, 
sensing data, for example, temperature, humidity, amount of chemicals, and soil conditions, 
have spatial correlation 
and correlation among data types in general \cite{Moltchanov}, \cite{Singh}.
The correlation information is often used in signal processing 
to improve system performance \cite{Gu}, \cite{Garg}.
The heterogeneous data correlation is introduced for the precoder design such as in \cite{Huh}. 
% or removed by whitening \cite{Xiao}.
However, the spatial correlation among sensors is ignored in many cases \cite{Liu, Zhu, Huh, Chen} 
or eliminated within the calculation process \cite{Cao}. 
% despite the inability to whiten at each sensor.

Motivated by the fact, 
we propose a novel precoder design for AirComp in wireless data aggregation 
that introduces correlation among sensors and data types, 
i.e., both the spatial correlation and heterogeneous data correlation.
We construct an optimization problem for designing the precoder 
by explicitly using the correlation.
This method can be applied to general cases 
where each sensor transmits a vector value, 
which is not considered in conventional methods such as \cite{WLiu, Liu, Cao}.
Moreover, 
we derive a closed-form solution without iterative procedures so that 
it achieves lower computational costs than conventional methods 
based on the iterative solutions of convex problems \cite{Huh, WLiu}.

In addition, this paper deals with a possible scenario in typical sensor networks.
Sensing devices usually have tiny batteries 
so that it is important 
to reduce power consumption related to wireless communication.
The conventional methods \cite{Chen} and \cite{Zhu} require square or vertically long precoding matrices 
which leads to dimensionality expansion of data. 
This procedure requires high communication costs for the sensors.
On the other hand, 
the proposed precoder enables dimensionality reduction~\cite{Varshney} of sensing data, 
which can reduce communication costs.
We conduct computer experiments focused on a situation where 
dimensionality reduction is performed by the precoder.

% The contributions of this paper are summarized as follows: 
% \begin{itemize}
% 	\item The proposed precoder is derived by explicitly 
%             employing the correlation among sensors and data types. 
%             That correlation is not included in the conventional precoding methods for AirComp.
% 	\item The proposed method requires no iterative procedures. 
%             This is motivated by the idea in \cite{Scaglione} and \cite{Sampath}, 
%             which is proposed in a different context from AirComp. 
%             The computational cost is greatly less than the gradient descent based method \cite{Huh}.
% 	\item Dimensionality reduction is introduced to the proposed method. 
%             This helps lower communication costs and reduce battery consumption. 
%             The conventional non-iterative methods \cite{Chen}, \cite{Zhu} are not applicable to the dimensional reduction.
% 	\item Simulation results on synthetic data show the superiority of the proposed method 
%             over other non-iterative solutions 
%             in cases where the number of receive antennas at the aggregator is less than 
%             that of the total transmit antennas at the sensors.
%             This indicates that the proposed method is especially suitable for large-scale sensor networks.
% \end{itemize}

\section{Preliminalies}
\subsection{Notations}
In the rest of the paper, we use the following notation.
Superscripts $(\cdot)^\mathrm{T}$ and $(\cdot)^{\mathrm{H}}$ denote the transpose and the Hermitian transpose, respectively.
The zero vector, zero matrix and identity matrix are represented as $\bm{0}$, $\bm{O}$, and $\bm{I}$, respectively.
$\ell_2$-norm is $\|\cdot\|$.
The complex circularly symmetric Gaussian distribution $\mathcal{CN}(\bm{0},\bm{\Sigma})$ has mean vector $\bm{0}$ and covariance matrix $\bm{\Sigma}$.
The expectation and trace operators are $\mathbb{E}[\cdot]$ and $\mathrm{Tr}[\cdot]$, respectively.
% The diagonal matrix is given by $\mathrm{diag}[\ldots]$ with the diagonal elements shown in the square brackets.
We denote the set of complex block diagonal matrices with $k$ diagonal blocks of $m\times n$ matrices as $\mathbb{B}_k^{m\times n}$.
% The $(i,j)$th element of a matrix is $[\cdot]_{ij}$.
The function $(\alpha)^+$ for $\alpha\in\mathbb{R}$ denotes $\max(0,\alpha)$.
Hadamard product is represented as $\otimes$, 
which is the elementwise multiplication of matrices.

\subsection{System Model}
\begin{figure}[tbp]
	\centerline{\includegraphics[width=\columnwidth]{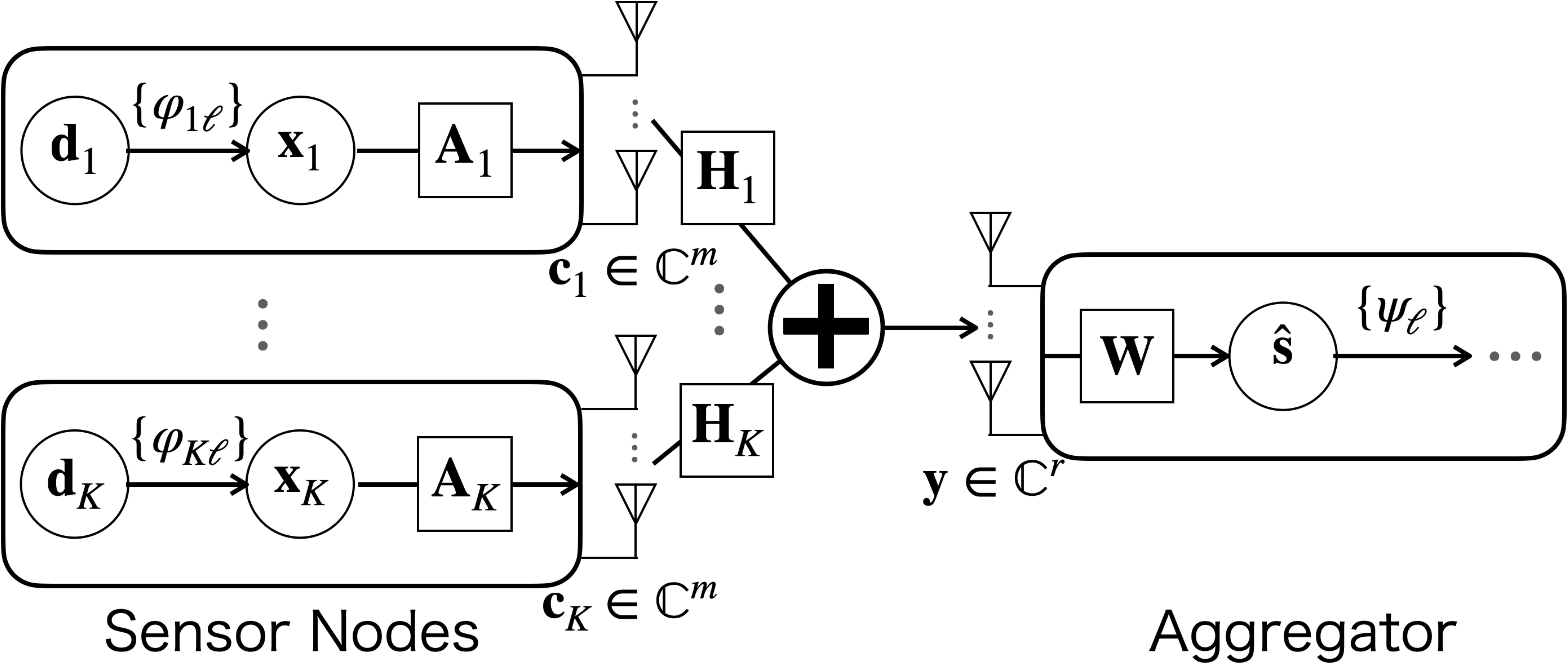}}
	\caption{System model of AirComp.}
	\label{fig:system}
\end{figure}
Assume a wireless data aggregation system with a single aggregator with $r$ receive antennas 
and $K$ sensor nodes with $m$ transmit antennas per node as illustrated in Fig.~\ref{fig:system}.

Let $\bm{d}_k\in\mathbb{R}^{n} \ (k=1,\ldots,K)$ be a vector composed of $n$ measurements at $k$th node 
and $d_{k\ell}\in\mathbb{R} \ (\ell=1,\ldots,n)$ be $\ell$th element of the vector.
The size should be set to $m< n$ to apply dimensionality reduction, 
but the following method is not limited to this setting.

In many applications of sensor networks,  
the objective of the aggregator is to obtain some function value of sensors' raw measurements. 
For example, arithmetic mean, weighted sum, or Euclidean norm is used as the function.
Such functions can be represented by combination of pre- and post-processing functions of the measurements and have been named nomographic functions \cite{Goldenbaum}, \cite{Zhu}.
The nomographic function $f_\ell(\cdot):\mathbb{R}\to\mathbb{R}$ is applied to each element of the vectors and given by 
\begin{equation}
    f_\ell(d_{1\ell},\ldots,d_{K\ell}) = \psi_\ell\left(\sum_{k=1}^K \varphi_{k\ell}(d_{k\ell})\right), 
    \label{eq:nom}
\end{equation}
where $\psi_\ell(\cdot):\mathbb{R}\to\mathbb{R}$ is a post-processing function 
and $\varphi_{k\ell}(\cdot):\mathbb{R}\to\mathbb{R}$ is a pre-processing function, respectively.
For example, the elementwise weighted sum 
$f_\ell(d_{1\ell},\ldots,d_{K\ell})=\sum_{k=1}^K \omega_{k\ell}d_{k\ell}$ 
can be represented by the pre-processsing function $\varphi_{k\ell}(\chi)=\omega_{k\ell}\chi$ 
and the post-processing one $\psi_\ell(\chi)=\chi$.

From the form of the nomographic function \eqref{eq:nom}, 
the aggregator wants to know the sum $\sum_{k=1}^K \varphi_{k\ell}(d_{k\ell})$ 
via communication with the nodes.
% In this paper, we consider employing AirComp to obtain the sum with low latency.
We redefine the pre-processed local function values at node $k$ as a vector 
$\bm{x}_k = [\varphi_{k1}(d_{k1}),\ldots,\varphi_{kn}(d_{kn})]^\mathrm{T}\in\mathbb{R}^n$ 
and then the elementwise sum can be summarized as 
$\bm{s}=\sum_{k=1}^K \bm{x}_k\in\mathbb{R}^n$.
Assume that the further summarized vector 
$\bm{x}=[\bm{x}_1^\mathrm{T},\ldots,\bm{x}_K^\mathrm{T}]^\mathrm{T}\in\mathbb{R}^{nK}$ 
follows $\bm{x}\sim\mathcal{CN}(\bm{0},\bm{K})$. 
The covariance matrix $\bm{K}=\mathbb{E}[\bm{xx}^\mathrm{T}]\in\mathbb{R}^{nK\times nK}$ is positive definite 
and known to the aggregator.
It includes information on the spatial correlation as the non-block diagonal elements 
and on the heterogeneous data correlation as the non-diagonal elements of each block.

Each node multiplies a linear precoder $\bm{A}_k\in\mathbb{C}^{m\times n}$ 
by its own pre-processed vector $\bm{x}_k$ for reducing aggregation error, 
that is, the node $k$ transmits 
\begin{equation}
    \bm{c}_k=\bm{A}_k\bm{x}_k \in \mathbb{C}^m 
\end{equation}
to the aggregator.
The size $m$ is assumed to be smaller than that of $\bm{x}_k$ for dimensionality reduction.
The aggregated signal through MAC is given by 
\begin{align}
	\bm{y} &= \bm{H}_1\bm{c}_1 + \cdots + \bm{H}_K\bm{c}_K + \bm{n} \nonumber\\
	&= \left(\sum_{k=1}^K \bm{H}_k\bm{A}_k\bm{x}_k\right) + \bm{n} \in \mathbb{C}^r,
    \label{eq:model1}
\end{align}
where $\bm{H}_k\in\mathbb{C}^{r\times m}$ is a channel matrix between node $k$ and the aggregator, 
and $\bm{n}\in\mathbb{C}^r$ is the additive noise that follows $\bm{n}\sim\mathcal{CN}(\bm{0},\bm{S})$.
The positive definite covariance matrix $\bm{S}\in\mathbb{C}^{r\times r}$ 
represents correlation of the noise vector and 
is known to the aggregator.
The model \eqref{eq:model1} is summarized as 
\begin{align}
    \bm{y} = \bm{H}\bm{A}\bm{x}+\bm{n},
\end{align}
where 
\[\bm{H}=\begin{bmatrix}\bm{H}_1, \ldots, \bm{H}_K \end{bmatrix}\in\mathbb{C}^{r\times mK}\] 
and 
\[\bm{A}=\begin{bmatrix} \bm{A}_1 & & \bm{O} \\ & \ddots & \\ \bm{O} & & \bm{A}_K \end{bmatrix}\in\mathbb{C}^{mK\times nK}.\]

The aggregator is assumed to know information of 
the statistical properties of the transmit signal $\bm{x}$ and the noise $\bm{n}$, 
the received signal $\bm{y}$, and the channel matrices $\{\bm{H}_k\}_{k=1}^K$.

\section{Proposed Method}
In this section, we describe how to design the precoder $\bm{A}$ 
by a non-iterative procedure while including the correlation of data. 

\subsection{Optimization Problem for Design of Proposed Precoder}
The objective at the aggregator is to obtain the sum 
\begin{equation}
    \bm{s}=\sum_{k=1}^K \bm{x}_k = \bm{Qx}, 
\end{equation}
where $\bm{Q}=[\bm{I},\ldots,\bm{I}]\in\mathbb{R}^{n\times nK}$, 
as correctly as possible by using the available information 
because the received signal is distorted by the channel and noise.
For this objective, 
we assume that the aggregator employs a linear MMSE estimate~\cite{MIMO} 
\begin{equation}
    \hat{\bm{s}} = \bm{Wy} = \bm{Q}\bm{KA}^\mathrm{H}\bm{H}^\mathrm{H}\left(\bm{HAKA}^\mathrm{H}\bm{H}^\mathrm{H}+\bm{S}\right)^{-1}\bm{y}. 
    \label{eq:mmse}
\end{equation}
The matrix 
\begin{equation}
    \bm{W}=\bm{Q}\bm{KA}^\mathrm{H}\bm{H}^\mathrm{H}\left(\bm{HAKA}^\mathrm{H}\bm{H}^\mathrm{H}+\bm{S}\right)^{-1}\in\mathbb{C}^{n\times r}
\end{equation} 
is determined 
by minimizing the MSE: 
\begin{equation}
    \mathbb{E}[\|\hat{\bm{s}}-\bm{s}\|^2] = \mathbb{E}\left[\left\|\bm{WHAx}+\bm{Wn}-\bm{Qx}\right\|^2\right].
\end{equation}

In this paper, 
we explore a matrix $\bm{A}$ that minimizes $\mathbb{E}[\|\hat{\bm{s}}-\bm{s}\|^2]$ 
of the linear MMSE estimation.
Moreover, the limited power of the sensor nodes should be also taken into consideration 
for the design of the matrix $\bm{A}$.
The total transmit power of the nodes is $\sum_{k=1}^K \mathbb{E}[\|\bm{A}_k\bm{x}_k\|^2] = \mathbb{E}[\|\bm{Ax}\|^2]$.
Therefore, we consider the optimization problem: 
\begin{align}
	(\mathcal{P}1) \quad \min_{\bm{A}}\ & \mathbb{E}\left[\|\hat{\bm{s}}-\bm{s}\|^2\right] \nonumber \\
	\mathrm{s.t.}\ &\mathbb{E}[\|\bm{Ax}\|^2] = P_0, \ \bm{A}\in\mathbb{B}_K^{m\times n},
\end{align}
where the first constraint means that the total transmit power is set to be $P_0(>0)$ 
and the second constraint means that $\bm{A}$ has a block diagonal structure.
The problem $\mathcal{P}1$ can be rewritten by substituting \eqref{eq:mmse} into the cost function 
and results in an optimization problem of matrix function:
\begin{align}
	(\mathcal{P}2) \quad \min_{\bm{A}}\ & \mathrm{Tr}\left[\bm{Q}\left(\bm{K}^{-1}+\bm{A}^\mathrm{H}\bm{H}^\mathrm{H}\bm{S}^{-1}\bm{HA}\right)^{-1}\bm{Q}^\mathrm{T}\right] \nonumber \\
	\mathrm{s.t.}\ & \mathrm{Tr}[\bm{AKA}^\mathrm{H}] = P_0, \ \bm{A}\in\mathbb{B}_K^{m\times n}.
\end{align}
The problem is nonconvex and difficult to solve in general.
To make matters worse, the matrix $\bm{A}$ to be optimized has a block diagonal structure, 
which complicates the optimization process.

\subsection{Policy}
This paper proposes a closed-form solution of the nonconvex problem $\mathcal{P}2$ 
along with the idea in the conventional methods \cite{Scaglione, Sampath} proposed in a different context from AirComp.
We first consider the problem: 
\begin{align}
	(\mathcal{P}2') \quad \min_{\bm{A}}\ & \mathrm{Tr}\left[\bm{Q}\left(\bm{K}^{-1}+\bm{A}^\mathrm{H}\bm{H}^\mathrm{H}\bm{S}^{-1}\bm{HA}\right)^{-1}\bm{Q}^\mathrm{T}\right] \nonumber \\
	\mathrm{s.t.}\ & \mathrm{Tr}[\bm{AKA}^\mathrm{H}] = P_0. 
\end{align}
This is a relaxed problem of $\mathcal{P}2$ 
where the block diagonal constraint is omitted. 
The problem $\mathcal{P}2'$ is known to have an optimal solution 
when the number of nodes is $K=1$ \cite{Scaglione, Sampath}.
We obtain the solution $\tilde{\bm{A}}$ of the problem $\mathcal{P}2'$ 
by using diagonalization of some matrices as employed in \cite{Scaglione} and \cite{Sampath}.
Next, a block diagonal matrix $\tilde{\bm{A}}_{\mathrm{bd}}$ is derived 
by removing non-block diagonal elements from $\tilde{\bm{A}}$.
This approach of the block diagonalization has been employed in \cite{Huang} for not-AirComp settings.
Finally, we obtain the solution $\hat{\bm{A}}$ 
by scalling the norm of $\tilde{\bm{A}}_{\mathrm{bd}}$ 
to satisfy the power constraint $\mathrm{Tr}[\hat{\bm{A}}\bm{K}\hat{\bm{A}}^\mathrm{H}] = P_0$.

The proposed procedure causes a loss of optimality when $K>1$.
However, it gives us a closed formula of the precoding matrix 
and allows derivation in lower computational costs than the related work \cite{Huh}.
We show the validity of the proposed method via experimental verification in Sect.~\ref{sec:sim1}.

\subsection{Derivation of Non-Blockdiagonal matrix $\tilde{\bm{A}}$}
This section derives non-block diagonal solution $\tilde{\bm{A}}$ 
by diagonalization of matrices in the objective function.
We describe eigenvalue decomposition of the matrices $\bm{K}$ and $\bm{H}^\mathrm{H}\bm{S}^{-1}\bm{H}$ as 
\begin{align}
	\bm{K}&=\bm{U\Delta U}^\mathrm{T}  \in \mathbb{R}^{nK\times nK},
	\label{eq:Kdecompose}\\
	\bm{H}^\mathrm{H}\bm{S}^{-1}\bm{H} &= \bm{V\Lambda V}^\mathrm{H} \in \mathbb{C}^{mK\times mK}, 
	\label{eq:HSHdecompose} 
\end{align}
respectively, where 
$\bm{U}\in\mathbb{R}^{nK\times nK}, \ \bm{V}\in\mathbb{C}^{mK\times mK}$ are unitary matrices, 
and $\bm{\Delta}\in\mathbb{R}_+^{nK\times nK}, \ \bm{\Lambda}\in\mathbb{R}_+^{mK\times mK}$ are diagonal matrices.
The diagonal elements are represented as 
\[\bm{\Delta} = \begin{bmatrix}
	\delta_1 &  & \bm{O} \\
	  & \ddots &  \\
	\bm{O} &  & \delta_{nK}
\end{bmatrix}, \ \bm{\Lambda} = \begin{bmatrix}
	\lambda_1 &  & \bm{O} \\
	  & \ddots &  \\
	\bm{O} &  & \lambda_{mK}
\end{bmatrix},\]
respectively.
Without loss of generality, we assume 
$\delta_1\geq\delta_2\geq\ldots\geq\delta_{nK}\geq0$ and 
$\lambda_1\geq\lambda_2\geq\ldots\geq\lambda_{mK}\geq0$.
When $r<mK$, the matrix $\bm{\Lambda}$ has the property $\lambda_{r+1}=\ldots=\lambda_{mK}=0$.
We further assume that the matrix variable $\bm{A}$ is decomposed as 
\begin{equation}
	\bm{A}=\bm{V\Phi U}^\mathrm{T}, \ 
	\bm{\Phi} = \begin{bmatrix}
		\phi_1 &  & \bm{O} &  & \\
		 & \ddots &  & \bm{O} & \\
		\bm{O} &  & \phi_{mK} &  & 
	\end{bmatrix}, 
	\label{eq:A}
\end{equation}
where $\{\phi_{j}\}_{j=1}^{mK}$ are scalar parameters and satisfy $|\phi_j|^2\geq0$.
That decomposed formulation is motivated by the work \cite{Scaglione, Sampath}.
Therefore, the solution of the problem $\mathcal{P}2'$ is assumed to 
be fully determined by the $mK$ parameters.

From these diagonalized representations, 
we can solve the problem $\mathcal{P}2'$ in terms of the parameters $\{\phi_{j}\}_{j=1}^{mK}$ 
as summarized in Theorem~1.
\begin{theorem}
	The solution of the problem $\mathcal{P}2'$ is given as the following water-filling problem 
	\begin{equation}
		|\hat{\phi}_j|^2 = \begin{cases}
			\frac{1}{\delta_j\lambda_j}\left(\sqrt{\frac{\delta_j\lambda_jR_j}{\mu}}-1\right)^+, \ (j=1,\ldots,\min(r,mK)), \\
			0, \ \quad\quad\quad\quad\quad\quad (j=\min(r,mK)+1,\ldots,mK),
		\end{cases}
        \label{eq:phijopt}
	\end{equation}
	where $R_j=\sum_{i=1}^n ([\bm{QU}]_{ij})^2$, 
    $[\bm{QU}]_{ij}$ is the $(i,j)$th element of $\bm{QU}$, 
	and $\mu\in\mathbb{R}$ is determined to satisfy the power constraint in the problem.
	For the case of $j=1,\ldots,\min(r,mK)$, 
	if the elements in $(\cdot)^+$ in the right-hand side of \eqref{eq:phijopt} are nonnegative for all $j$, 
	the solution is given by 
	\begin{align}
		\hat{\phi}_j &= 
		\sqrt{\frac{1}{\delta_j\lambda_j}\left(\sqrt{\delta_j\lambda_jR_j}
		\frac{P_0+\sum_{l=1}^{\min(r,mK)}\frac{1}{\lambda_l}}
		{\sum_{l=1}^{\min(r,mK)}\sqrt{\frac{\delta_lR_l}{\lambda_l}}}-1\right)}.
		\label{eq:optphij}
	\end{align}
\end{theorem}
\textit{Proof}: 
We represent the cost function of the problem $\mathcal{P}2'$ as $f(\{\phi_j\}_{j=1}^{mK})$.
The function can be rewritten as 
\begin{align}
	f(\{\phi_j\}_{j=1}^{mK})
	&=\mathrm{Tr}\left[\bm{QU}\left(\bm{\Delta}^{-1}+\bm{\Phi}^\mathrm{H}\bm{\Lambda \Phi}\right)^{-1}\bm{U}^\mathrm{T}\bm{Q}^\mathrm{T}\right]
	\label{eq:tmpmse}
\end{align}
by using \eqref{eq:Kdecompose}--\eqref{eq:A}.
Note that the matrices $\bm{Q}$ and $\bm{U}$ are known at the aggregator 
and then it is possible to calculate the trace in \eqref{eq:tmpmse} directly.
The result of expanding the equation is 
\begin{align}
	f(\{\phi_j\}_{j=1}^{mK})
	&=\sum_{j=1}^{\min(r,mK)}\frac{\delta_j R_j}{1+\delta_j \lambda_j |\phi_j|^2} 
	+ \sum_{j=\min(r,mK)+1}^{nK}\delta_j R_j.
\end{align}
On the other hand, the transmit power $\mathrm{Tr}[\bm{AKA}^\mathrm{H}]$ in the problem $\mathcal{P}2'$ 
is also given by 
\begin{equation}
	\mathrm{Tr}[\bm{AKA}^\mathrm{H}] = \mathrm{Tr}[\bm{\Phi\Delta\Phi}^\mathrm{H}]
	= \sum_{j=1}^{mK} \delta_j|\phi_j|^2.
\end{equation}

In order to obtain the solution of the problem $\mathcal{P}2'$, 
we set the following Lagrangian function 
\begin{equation}
	\mathcal{L}(\{\phi_j\}_{j=1}^{mK},\mu) 
	= f(\{\phi_j\}_{j=1}^{mK})
	- \mu\left(P_0-\sum_{j=1}^{mK} \delta_j|\phi_j|^2\right), 
	\label{eq:lag}
\end{equation}
where $\mu$ is the Lagrange multiplier.
% The Karush-Kuhn-Tucker (KKT) conditions in this case are 
The condition in this case is $\frac{\partial \mathcal{L}}{\partial |\phi_j|^2}=\frac{\partial \mathcal{L}}{\partial \mu}=0$.
% $\frac{\partial \mathcal{L}}{\partial |\phi_j|^2}=\frac{\partial \mathcal{L}}{\partial \mu}=0$, $\mu\geq0$, 
% $P_0-\sum_{j=1}^{mK} \delta_j|\phi_j|^2\leq 0$, 
% and $\mu\left(P_0-\sum_{j=1}^{mK} \delta_j|\phi_j|^2\right) = 0$.
By solving $\frac{\partial \mathcal{L}}{\partial |\phi_j|^2}=0$ 
in terms of $|\phi_j|^2$, 
we can obtain 
\begin{equation}
	|\phi_j|^2 = \begin{cases}
		\frac{1}{\delta_j\lambda_j}\left(\sqrt{\frac{\delta_j\lambda_jR_j}{\mu}}-1\right)^+, \ (j=1,\ldots,\min(r,mK)), \\
		0, \ \quad\quad\quad\quad\quad\quad (j=\min(r,mK)+1,\ldots,mK), 
	\end{cases}
	\label{eq:phij}
\end{equation}
% where the function $(\cdot)^+$ is applied because of the KKT conditions.
where the function $(\cdot)^+$ is applied because $|\phi_j|^2\geq0$.

If the arguments in the right-hand side of \eqref{eq:phij} become nonnegative for all indeces $j$, 
another relation can be derived by substituting \eqref{eq:phij} 
into $\frac{\partial \mathcal{L}}{\partial \mu}=0$ and we then have 
\begin{equation}
	\frac{1}{\sqrt{\mu}} = 
	\frac{P_0+\sum_{j=1}^{\min(r,mK)}\frac{1}{\lambda_j}}
	{\sum_{j=1}^{\min(r,mK)}\sqrt{\frac{\delta_jR_j}{\lambda_j}}}.
	\label{eq:mu}
\end{equation}
From \eqref{eq:phij} and \eqref{eq:mu}, 
we can obtain the solution $\hat{\phi}_j$.

If there exists index $j$ 
where the argument in the right-hand side of \eqref{eq:phij} becomes negative, 
the Lagrange multiplier $\mu$ is determined by the well-known water-filling algorithm (Sect.~3.E in \cite{Palomarbook}) 
to satisfy the power constraint.
\qed

We can obtain the matrix $\tilde{\bm{A}}$ 
by using the solution $\{\hat{\phi}_j\}_{j=1}^{mK}$ 
and constructing from \eqref{eq:A}.

\subsection{Block Diagonalization}
\label{sec:blkdiag}
The problem we should solve is $\mathcal{P}2$ 
and the matrix $\bm{A}$ must have block-diagonal structure.
In this paper, 
we omit the non-block diagonal elements of the matrix $\tilde{\bm{A}}$ obtained in the previous section 
and then rescale to the constrained power.

Let $\bm{M}\in\mathbb{B}_K^{m\times n}$ be a masking matrix composed of $K\times K$ blocks 
where the diagonal blocks are the matrices whose components are all $1$ 
and the non-diagonal blocks are all $\bm{O}$.
The block diagonalized matrix $\tilde{\bm{A}}_\mathrm{bd}$ can be represented as 
the elementwise multiplication of $\bm{M}$ and $\tilde{\bm{A}}$, 
i.e., 
\begin{equation}
	\tilde{\bm{A}}_\mathrm{bd} = \bm{M}\otimes\tilde{\bm{A}}.
\end{equation}
Note that the matrix $\tilde{\bm{A}}$ satisfies the power constraint 
$\mathrm{Tr}[\tilde{\bm{A}}\bm{K}\tilde{\bm{A}}^\mathrm{H}]=P_0$ 
because of the constraint of the problem $\mathcal{P}2'$ but 
the block diagonalized matrix $\tilde{\bm{A}}_\mathrm{bd}$ does not.
Therefore, we rescale the norm of the matrix $\tilde{\bm{A}}_\mathrm{bd}$ to satisfy the power constraint.
We then obtain the final solution 
\begin{equation}
	\hat{\bm{A}} = \sqrt{\frac{P_0}{\mathrm{Tr}[\tilde{\bm{A}}_\mathrm{bd}\bm{K}\tilde{\bm{A}}_\mathrm{bd}^\mathrm{H}]}} \tilde{\bm{A}}_\mathrm{bd}.
\end{equation}

The proposed method does not include iterative processes 
and the main factor of the computational costs is eigenvalue decomposition \eqref{eq:Kdecompose}, 
which requires $\mathcal{O}((nK)^3)$.
However, this complexity is much less than the conventional iterative method \cite{Huh}.

% Note that, if there is only one node in a network, i.e., $K=1$, 
% the resulting precodor of the proposed method becomes optimal, 
% though the settings makes no sense in AirComp.
% The optimality has been shown in \cite{Scaglione, Sampath}.

\section{Simulation Results}
\label{sec:sim1}
Performance with the proposed precoder was evaluated via computer simulations.
We evaluated influence of system parameters 
on the averaged and normalized squared error, i.e., 
$\sum_{t=1}^T\sum_{z=1}^Z\|\hat{\bm{s}}_{tz}-\bm{s}_{tz}\|^2/(nKTZ)$, 
where $T$ is the number of generations of $\bm{H}$, 
$Z$ is the number of generations of $\bm{x}$ for a single generation of $\bm{H}$, 
and $\hat{\bm{s}}_{tz}$ and $\bm{s}_{tz}$ are corresponding instances.
We set to $T=10$ or more and $Z=100$.
Specifically, the simulation results are examined in terms of 
the performance dependency on 
\begin{itemize}
	\item Data compression ratio $mK/nK$: ratio of the total number of transmit antennas and that of measurements, 
	\item Communication compression ratio $r/mK$: ratio of the number of receive antennas and the total number of transmit antennas, 
	\item The number of nodes $K$, 
	\item and Signal-to-noise ratio (SNR) (dB).
\end{itemize}
In this paper, we define the SNR as $10\log_{10}(P_0/\mathrm{Tr}[\bm{S}])$(dB).
The length of the measurement was set to $n=8$ and 
the total transmit power was $P_0=10$ in all the simulations.
The covariance matrices $\bm{K}$ and $\bm{S}$ had correlated formulations 
and the elements were determined as 
\[[\bm{K}]_{ij}=0.8^{|i-j|}, \ [\bm{S}]_{ab}=0.5^{|a-b|}/r,\]
for $i,j=1,\ldots,nK$ and $a,b=1,\ldots,r$, respectively.
The instances $\bm{x}$ and $\bm{n}$ were randomly generated 
from the distributions $\mathcal{CN}(\bm{0},\bm{K})$ and $\mathcal{CN}(\bm{0},\bm{S})$, respectively.
Each element of the channel matrix $\bm{H}$ was 
identically and independently generated by $\mathcal{CN}(0,1)$.
Moreover, we compared the performance of the proposed method with 
the following four schemes: 
\begin{enumerate}
    \item ignoring correlation: the proposed method that ignores spatial correlation, namely, ignores non-block diagonal elements of $\bm{K}$ when designing the precoder, 
	\item Communicate-then-Compute: method not specifically designed for AirComp by using MSE regarding $\bm{x}$, 
    where the optimization is done in the same manner as \cite{Scaglione, Sampath} and then applied block diagonalization in Sect.~\ref{sec:blkdiag}, 
	\item Random: method using a random matrix as $\bm{A}$ where each element is identically and independently generated by $\mathcal{CN}(0,1)$ and normalized to satisfy the power constraint. Such a dimensional reduction is typically employed in some estimation methods \cite{Woodruff}, 
	\item and Huh et al.: method iteratively solving convex problems for deriving the matrices $\bm{A}_k$ from the problem $\mathcal{P}1$ \cite{Huh}. The method requires higher computational cost than the proposed and other methods so that it is regarded as a baseline.
\end{enumerate}
The number of iterations for the method \cite{Huh} was set to $10$.
% The performance of the gradient descent based method should be the best for every experiment 
% at the cost of computational complexity.
% This method was employed as a baseline.

\begin{figure}[tbp]
    % \centerline{\includegraphics[width=\columnwidth]{figures/20220428_n8ratior5k30P10snr25numitr10000freqH100numgrid10-3_vsm_normalize.pdf}}
    \centerline{\includegraphics[width=\columnwidth]{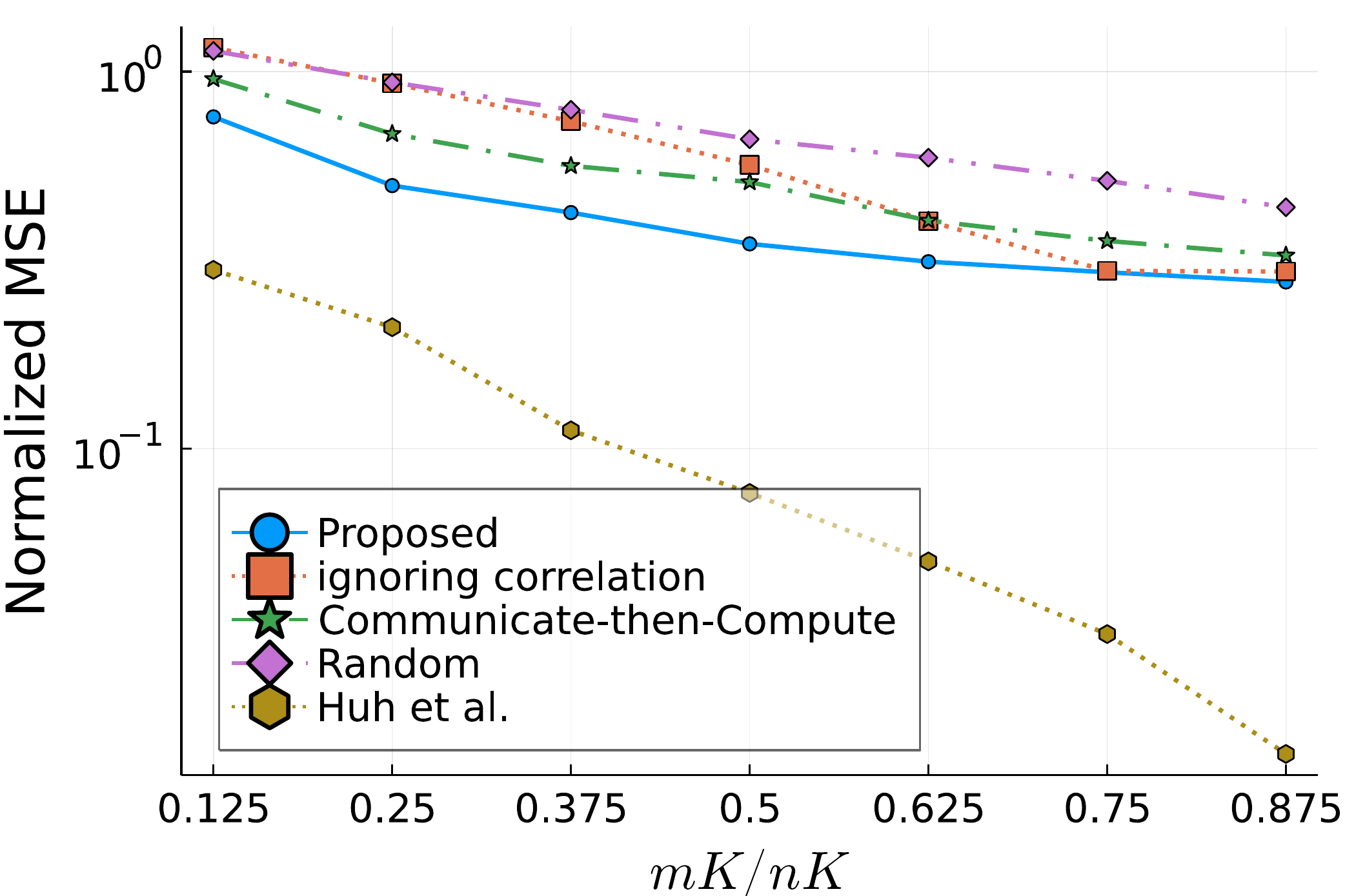}}
    \caption{MSE vs. data compression ratio where $(n,K)=(8,30)$, $r=5m$, and SNR$=25$(dB).}
    \label{fig:vsm2}
\end{figure}

Fig.~\ref{fig:vsm2} shows the evaluation with respect to the data compression ratio.
The system parameters were set to $(K,r,\mathrm{SNR(dB)})=(30,5m,25)$.
From the figure, the iterative method of Huh et al. achieves the lowest error at the cost of high computational costs.
The proposed method shows the best performance at any data compression ratio among the non-iterative methods.
% \begin{figure}[tbp]
%     \centerline{\includegraphics[width=\columnwidth]{figures/20220407_n4ratior5k6P10snr25numitr100000freqH100_mse_normalize.pdf}}
%     \caption{$(n,K)=(4,6)$, $r=5m$, and SNR$=25$(dB).}
%     \label{fig:vsm}
% \end{figure}

\begin{figure}[tbp]
    % \centerline{\includegraphics[width=\columnwidth]{figures/20220428_n8m2k30P10snr25numitr10000freqH100numgrids10-3_vsr_normalize.pdf}}
    \centerline{\includegraphics[width=\columnwidth]{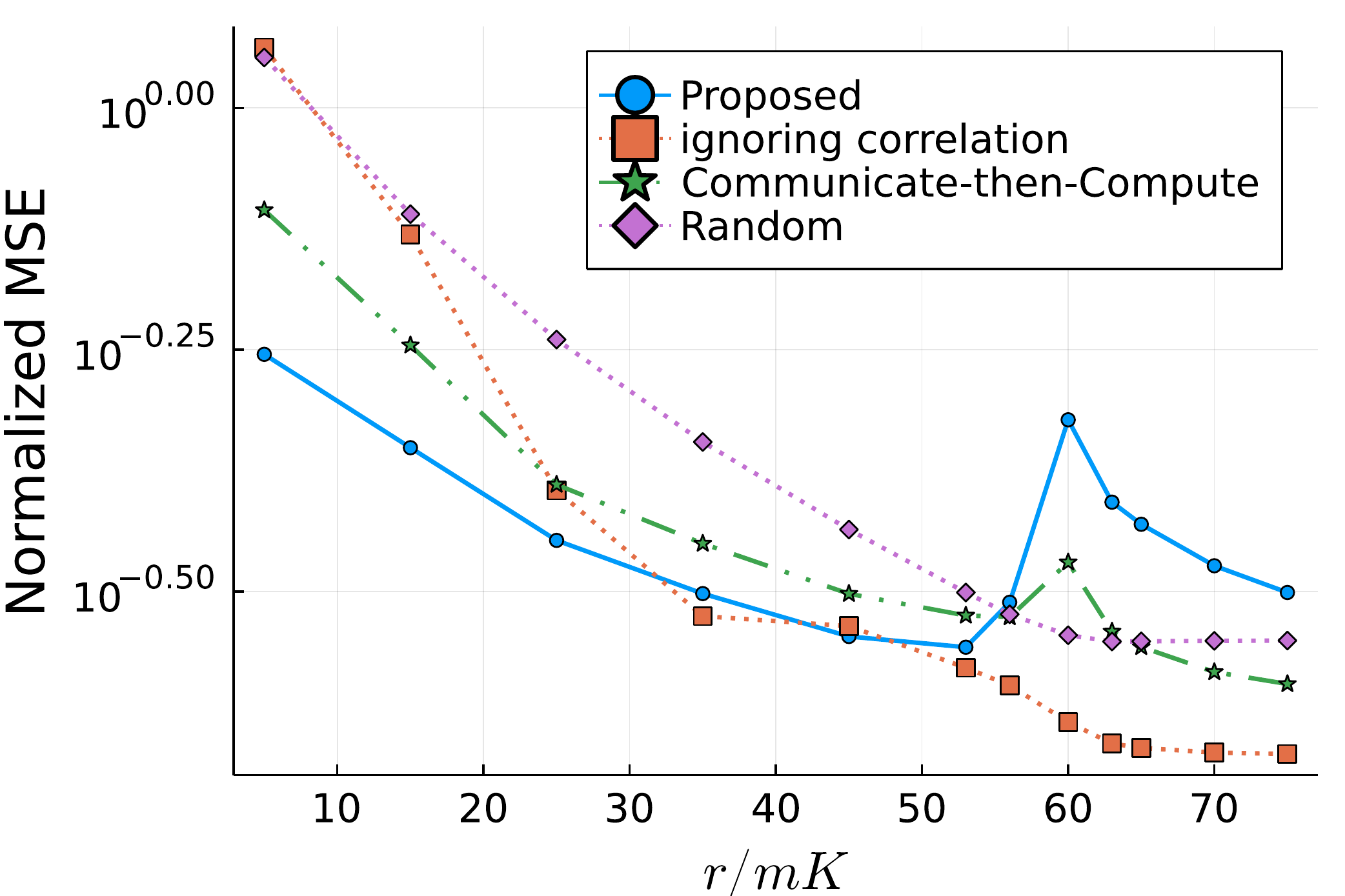}}
    \caption{MSE vs. communication compression ratio where $(n,m,K)=(8,2,30)$ and SNR$=25$(dB).}
    \label{fig:vsr2}
\end{figure}

\begin{figure}[tbp]
    % \centerline{\includegraphics[width=\columnwidth]{figures/20220428_n8m2r16P10snr25numitr10000freqH100numgrids10-3_vsk2_normalize.pdf}}
    \centerline{\includegraphics[width=\columnwidth]{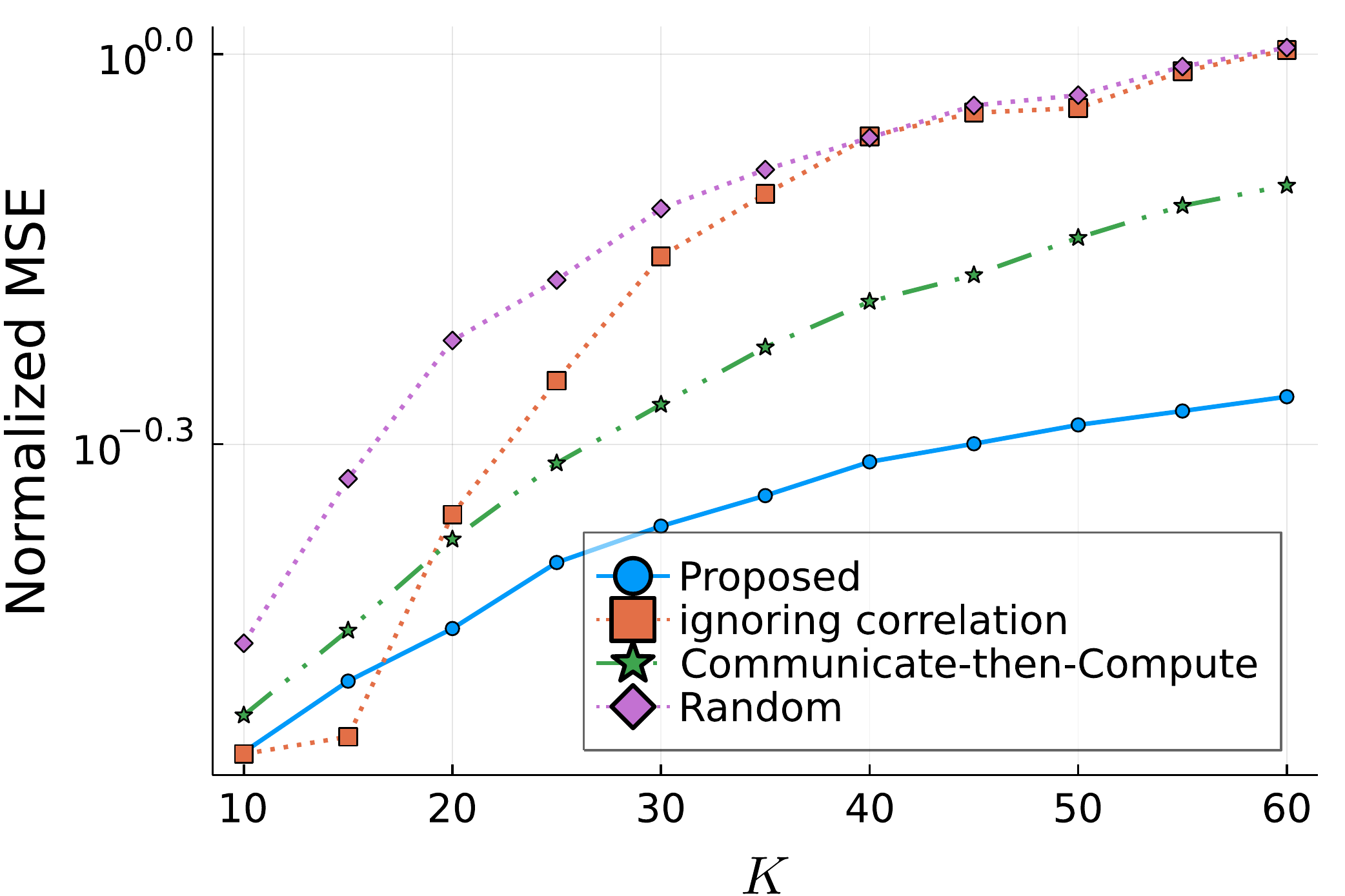}}
    \caption{MSE vs. the number of nodes where $(n,m,r)=(8,2,16)$ and SNR$=25$(dB).}
    \label{fig:vsk2}
\end{figure}

The key feature of the proposed method is revealed 
from viewpoints of communication compression ratio and the number of nodes, 
shown in Fig.~\ref{fig:vsr2} and Fig.~\ref{fig:vsk2}, respectively.

The evaluation in terms of communication compression ratio is shown in Fig.~\ref{fig:vsr2}, 
where the system parameters were set to $(m,K,\mathrm{SNR(dB)})=(2,30,25)$.
The MSE curves of the proposed and communicate-then-compute methods 
appear to have two modes.
In the region $1\leq r/mK$, 
i.e., when the number of the receive antennas $r$ is sufficiently large, 
the proposed method shows higher error than the other methods.
This may cause by the suboptimality of the proposed method.
On the other hand, in the region $r/mK<1$, 
the proposed method achieves lower errors than the other methods in most cases.

In Fig.~\ref{fig:vsk2}, we evaluated the influence of the number $K$ of nodes on normalized MSE.
The system parameters were set to $(m,r,\mathrm{SNR(dB)})=(2,16,25)$.
In all the methods, 
the smaller the number of nodes, the lower the estimation error.
The error of the proposed method is the lowest when $K\geq 20$ and 
the performance difference from the other methods becomes larger with increase of $K$.
% Note that, when $K=1$, the proposed and the communicate-then-compute method 
% are identical and the solutions has been shown to be optimal \cite{Scaglione, Sampath}.
% As with Fig.~\ref{fig:vsr2}, in the region $9\leq K$, 
% i.e., $r/mK=16/2K<1$, 
% the proposed method shows better performance.
Compared with the method ignoring correlation, 
we can see that the spatial correlation should be exploited for the precoder design 
especially when there are a large number of nodes in a network.

These results indicate that the proposed method is suitable 
in situations where the number of receive antennas are limited 
but the number of nodes is increasing. 
The situations are nothing short of typical IoT environments.
% \begin{figure}[tbp]
%     \centerline{\includegraphics[width=\columnwidth]{figures/20220329_n4m1k6P10snr25numitr10000freqH100numopt10_mse_normalize.pdf}}
%     \caption{$(n,m,K)=(4,1,6)$ and SNR$=25$(dB).}
%     \label{fig:vsr}
% \end{figure}
% \begin{figure}[tbp]
%     \centerline{\includegraphics[width=\columnwidth]{figures/20220407_n4m1r5P10snr25numitr100000freqH100numopt10_normalized.pdf}}
%     \caption{$(n,m,r)=(4,1,5)$ and SNR$=25$(dB).}
%     \label{fig:vsk}
% \end{figure}

\begin{figure}[tbp]
    % \centerline{\includegraphics[width=\columnwidth]{figures/20220428_n8m2r30k30P10numitr10000freqH100numgrids10-3_vsSNR_normalize.pdf}}
    \centerline{\includegraphics[width=\columnwidth]{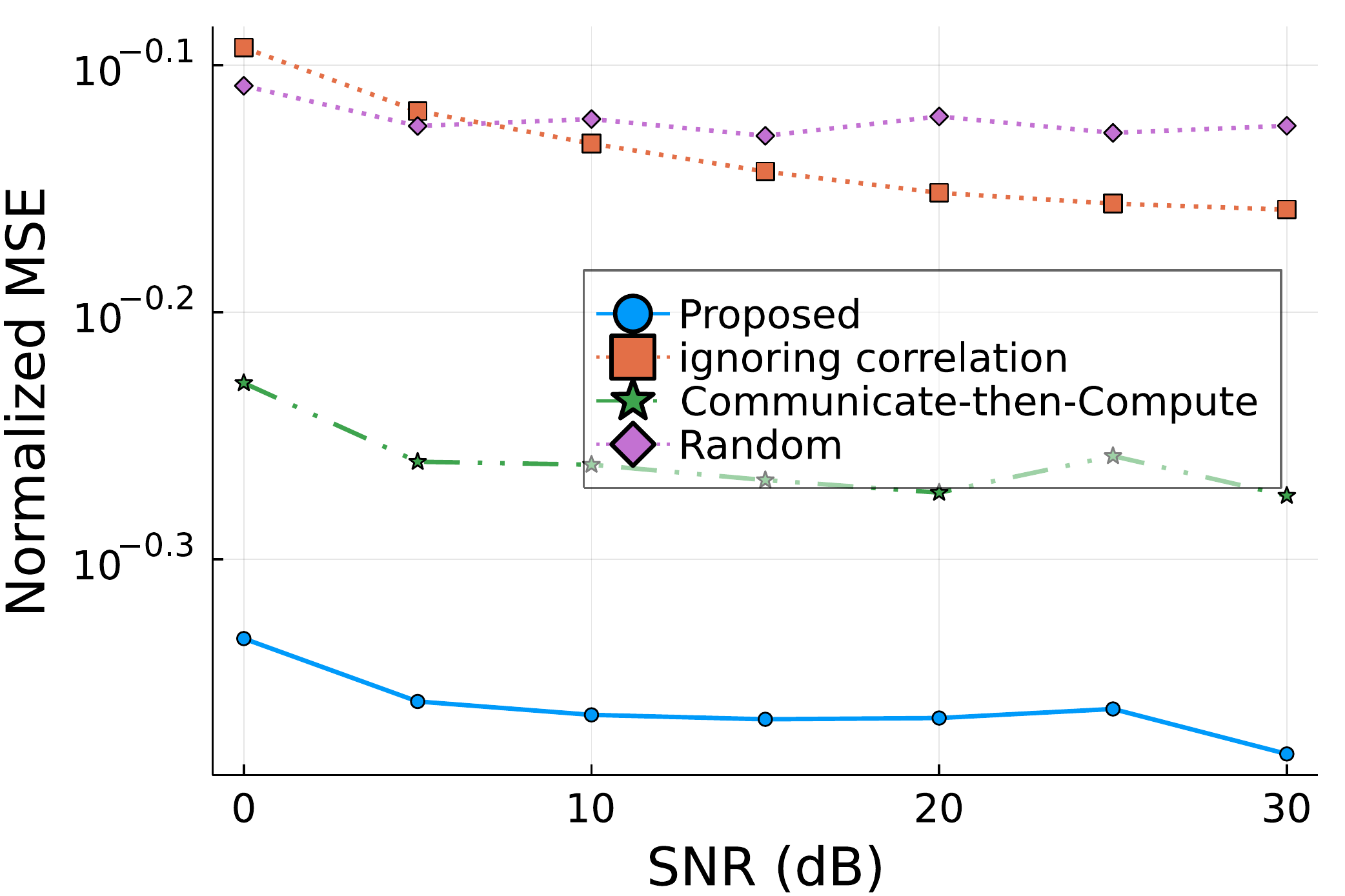}}
    \caption{MSE vs. SNR where $(n,m,r,K)=(8,2,16,30)$.}
    \label{fig:vssnr2}
\end{figure}

Finally, we evaluated performance of the proposed method in different SNR.
The system parameters were set to $(m,r,K)=(2,16,30)$.
From Fig.~\ref{fig:vssnr2}, the performance of the proposed method is the best 
among the methods at any SNR.
% \begin{figure}[tbp]
%     \centerline{\includegraphics[width=\columnwidth]{figures/20220329_n4m1r5k6P10numitr100000freqH100numopt10_mse_normalize.pdf}}
%     \caption{$(n,m,r,K)=(4,1,5,6)$.}
%     \label{fig:vssnr}
% \end{figure}

\section{Conclusions}
This paper has proposed a novel precoder design for Aircomp in wireless data aggregation 
that explicitly employs spatial correlation and heterogeneous data correlation. 
The correlation appears in typical applications of sensor networks such as environmental monitoring 
and the appropriate use helps reduce aggregation errors that occurred in AirComp.
The proposed method includes no iterative procedure so that it does not require high computational costs.
This is motivated by the idea of matrix diagonalization proposed in a different context from AirComp.
Furthermore, this method provides dimensionality reduction of the transmit vectors 
which helps reduce communication costs per sensor.

Simulation results on synthetic data showed that 
the performance of the proposed method including correlation is better than 
the method ignoring correlation and the other non-iterative methods, 
especially when there are a lot of nodes in a network and 
when the number of receive antennas at the aggregator is less 
than the total number of transmit antennas at the sensors. 
In other words, the proposed method achieves better performance in typical large-scale IoT environments.

Future work includes the extension of the proposed method 
involving more sophisticated operations of block diagonalization 
and simulation on real datasets.

% \begin{table}[htbp]
% \caption{Table Type Styles}
% \begin{center}
% \begin{tabular}{|c|c|c|c|}
% \hline
% \textbf{Table}&\multicolumn{3}{|c|}{\textbf{Table Column Head}} \\
% \cline{2-4} 
% \textbf{Head} & \textbf{\textit{Table column subhead}}& \textbf{\textit{Subhead}}& \textbf{\textit{Subhead}} \\
% \hline
% copy& More table copy$^{\mathrm{a}}$& &  \\
% \hline
% \multicolumn{4}{l}{$^{\mathrm{a}}$Sample of a Table footnote.}
% \end{tabular}
% \label{tab1}
% \end{center}
% \end{table}

% \begin{figure}[htbp]
% \centerline{\includegraphics{fig1.png}}
% \caption{Example of a figure caption.}
% \label{fig}
% \end{figure}

\end{document}